\documentclass[aps,prd,nofootinbib,amsmath,amssymb,showpacs,superscriptaddress,twocolumn]{revtex4}
\usepackage{txfonts}
\usepackage{graphicx}
\usepackage{dcolumn}
\usepackage{bm}
\usepackage{amssymb}
\usepackage{latexsym}
\usepackage[colorlinks, linkcolor=blue, citecolor=blue, urlcolor=blue]{hyperref}

\newcommand{\be}{\begin{equation}}
\newcommand{\ee}{\end{equation}}

\def\bea{\begin{eqnarray}}
\def\eea{\end{eqnarray}}

\begin{document}

\title{Probing the dynamics of dark energy with divergence-free parametrizations:
A global fit study}

\author{Hong Li}
\affiliation{Institute of High Energy Physics, Chinese Academy of
Sciences, Beijing 100049, China} \affiliation{Theoretical Physics
Center for Science Facilities, Chinese Academy of Sciences, Beijing
100049, China}
\author{Xin Zhang}
\affiliation{Department of Physics, College of Sciences,
Northeastern University, Shenyang 110004, China} \affiliation{Center
for High Energy Physics, Peking University, Beijing 100080, China}

\begin{abstract}
The CPL parametrization is very important for investigating the
property of dark energy with observational data. However, the CPL
parametrization only respects the past evolution of dark energy but
does not care about the future evolution of dark energy, since
$w(z)$ diverges in the distant future. In a recent paper [J.Z. Ma
and X. Zhang, Phys.\ Lett.\  B {\bf 699}, 233 (2011)], a robust,
novel parametrization for dark energy, $w(z)=w_0+w_1({\ln (2+z)\over
1+z}-\ln2)$, has been proposed, successfully avoiding the future
divergence problem in the CPL parametrization. On the other hand, an
oscillating parametrization (motivated by an oscillating quintom
model) can also avoid the future divergence problem. In this Letter,
we use the two divergence-free parametrizations to probe the
dynamics of dark energy in the whole evolutionary history. In light
of the data from 7-year WMAP temperature and polarization power
spectra, matter power spectrum of SDSS DR7, and SN Ia Union2 sample,
we perform a full Markov Chain Monte Carlo exploration for the two
dynamical dark energy models. We find that the best-fit dark energy
model is a quintom model with the EOS across $-1$ during the
evolution. However, though the quintom model is more favored, we
find that the cosmological constant still cannot be excluded.
\end{abstract}

\pacs{95.36.+x, 98.80.Es, 98.80.-k}

\keywords{Dark energy; divergence-free parametrizations; global fit;
Markov Chain Monte Carlo approach}

\maketitle


\section{Introduction}

Since the accelerating expansion of the universe was discovered by
the observations of type Ia supernovae (SN Ia) \cite{Riess98}, dark
energy, the mysterious energy budget that drives such a cosmic
acceleration, has attracted lots of studies \cite{dereview,de2}. The
main characteristic of dark energy is encoded in the equation of
state parameter (EOS), and thus the study of extracting the
information of EOS by fitting with the observational data provides
an important way for understanding the nature of dark energy.

Extracting the information of the EOS from the data relies on the
parametrization of dark energy. Currently, beyond the simplest
$\Lambda$CDM model, the Chevallier-Polarski-Linder (CPL, hereafter)
parametrization $w(z)= w_0 +w_a(1-a)$ \cite{CPL}, which introduces
the first order Taylor's expansion in terms of the scale factor $a$,
is rather popular and attracts lots of studies. The main feature of
such a parametrization is that it describes the possible dynamical
evolution of EOS with time. The advantage of this form is that it
can be applied to fit the low redshift SN Ia data as well as the
high redshit CMB data at the same time. However, as shown in our
previous study \cite{Ma:2011nc}, the EOS will get to a nonphysical
value in the far future time when redshift $z$ approaches $-1$,
namely, $|w(z)|$ will grow rapidly and diverge. Such a divergence
feature prevents the CPL parametrization from genuinely covering the
scalar-field models as well as other theoretical models.

The ultimate fate of the universe is determined by the property of
dark energy: If dark energy is the cosmological constant ($w=-1$),
then the fate of the universe is a de Sitter spacetime; if dark
energy is phantomlike ($w<-1$), then the destiny of the universe is
a doomsday (namely, the ``big rip'' singularity); and so on. So, it
is very important to probe the dynamics of dark energy with the
observational data, since the detection of the evolution of dark
energy would provide the evidence of falsification of the
cosmological constant, and also the ultimate fate of the universe
could be foreseen. Since the CPL parametrization has the divergence
problem and thus loses the prediction ability, we are interested in
some other well-behaved parametrization forms for investigating the
property of dark energy.

In order to keep the advantage of the CPL parametrization, and avoid
its drawback at the same time, some divergence-free
parameterizations have been proposed \cite{Ma:2011nc} in which the
leading proposal is a logarithm form: $w(z) = w_{0} +
w_{1}(\frac{\ln(2+z)}{1+z}-\ln2)$. Such a new parametrization has
well behaved, bounded behavior for both high redshifts and negative
redshifts. Thanks to the logarithm form in the parametrization, a
finite value for $w(z)$ can be ensured, via the application of the
L'Hospital's rule, in both limiting cases, $z\rightarrow\infty$ and
$z\rightarrow -1$. This is the reason why a logarithm form is
introduced in the new parametrization. Without doubt, such a
two-parameter form of EOS can genuinely cover many scalar-field
models (including quintom models with two scalar fields and/or one
scalar field with high derivatives) as well as other theoretical
scenarios. On the other hand, one can only justify that the EOS of
dark energy is around $-1$ in the recent epoch, but for the EOS, in
much earlier or far future time, there are more possibilities, and
one of which is that the EOS of dark energy might exhibit
oscillatory behavior during the evolution.  Oscillating EOS of dark
energy are widely studied
\cite{Dodelson:2001fq,Feng:2004ff,Xia:2004rw,Linder:2005dw,Zhao:2006qg},
thanks to the advantage that the oscillating evolution behavior of
dark energy can unify the two accelerating epochs of our universe
and alleviate the so-called coincidence problem in some sense. Based
on this consideration, the above new parametrization is also
extended to an oscillating form. In Ref.~\cite{Ma:2011nc}, two novel
parametrizations have been used to probe the dynamics of dark energy
in the whole evolutionary history, and it has been proven that the
divergence-free parametrizations are very successful in exploring
the dynamical evolution of dark energy and have powerful prediction
capability for the ultimate fate of the universe.

In this Letter, we perform a global data fitting analysis on two
divergence-free parametrizations for dynamical dark energy, and
present constraints on the model parameters from the current
observational data, including the CMB temperature and polarization
power spectra from the seven-year WMAP data, the matter power
spectrum from the SDSS Data Release 7 (DR7), and SN Union2 sample.
Since dark energy parameters are tightly correlated to some other
cosmological parameters, for example, the matter density parameter
$\Omega_m$, the Hubble constant $H_0$, the spatial curvature
$\Omega_k$, the neutrino mass $m_\nu$, and so on, it is crucial to
consider a global fit procedure in the investigation of the
dynamical dark energy. Also, in this procedure, the perturbation of
dark energy is involved. In Ref.~\cite{Ma:2011nc}, only a
preliminary analysis was performed, in which the perturbation of
dark energy is absent, and the information of CMB and LSS is
incomplete. In this Letter, we will perform a sophisticated analysis
for the divergence-free parametrizations. The Letter is organized as
follows: In Sec.~II we will introduce the method and data of the
global fitting procedure, and the results are presented in Sec.~III,
and our conclusion is given in Sec.~IV.


\section{Method and Data}
\label{Method}

We consider the divergence-free parametrization for dynamical dark
energy proposed in Ref.~\cite{Ma:2011nc}:
\begin{equation}\label{pmt1}
w(z) = w_{0} + w_{1}\left(\frac{\ln(2+z)}{1+z}-\ln2\right),
\end{equation}
where $w_0$ denotes the present-day value of $w(z)$, and $w_1$ is
another parameter characterizing the evolution of $w(z)$. Note that
a minus $\ln2$ in the bracket is kept for maintaining $w_0$ to be
the current value of $w(z)$, and in Ref.~\cite{Ma:2011nc} it is
contrived for an easy comparison with the CPL model. Obviously, this
new parametrization has well behaved, bounded behavior for both high
redshifts and negative redshifts. The logarithm form in the
parametrization ensures a finite value for $w(z)$ via the
application of the L'Hospital's rule, in both limiting cases,
$z\rightarrow\infty$ and $z\rightarrow -1$. This is the reason why
we introduce a logarithm form in this parametrization. Specifically,
we have $w=w_0-w_1\ln2$ for $z\rightarrow\infty$ and
$w=w_0+w_1(1-\ln2)$ for $z\rightarrow -1$. At low redshifts, this
parametrization form reduces to the linear one, $w(z)\approx
w_0+\tilde{w}_1z$, where $\tilde{w}_1=-(\ln2)w_1$. Of course, one
can also recast it at low redshifts as the CPL form, $w(z)\approx
w_0+\tilde{w}_1z/(1+z)$, where $\tilde{w}_1=(1/2-\ln2)w_1$.
Therefore, it is clear to see that this parametrization exhibits
well-behaved feature for the dynamical evolution of dark energy.
Without question, such a two-parameter form of EOS can genuinely
cover scalar-field models as well as other theoretical scenarios.

The oscillating parametrization proposed in Ref.~\cite{Ma:2011nc} is
of the form $w(z)=w_0+w_1(\sin(1+z)/(1+z)-\sin(1))$. This form has
lots of advantages, as illustrated in Ref.~\cite{Ma:2011nc}. We find
that this parametrization describes the same behavior as the
logarithm form (\ref{pmt1}) at low redshifts, but exhibits
oscillating feature from a long term point of view. However, recent
studies show that there might be oscillatory behavior within
redshift range from 0 to 2 for the EOS \cite{Wu:2010av}, so this
possibility should also be involved in our investigation. Therefore,
we adopt the following oscillatory parametrization form:
\begin{equation}\label{pmt2}
w(z) = w_{0} + w_{1}\sin(A\ln(1/(1+z))),
\end{equation}
where $A$ is another parameter. The direct physical motivation of
the parametrization (\ref{pmt2}) is from an oscillating quintom
\cite{Feng:2004ff,Xia:2004rw}. Here a sine function has the
advantage of exhibiting the oscillating feature of the EOS and
preserving the value of EOS finite. In this Letter we set $A$ to be
$\frac{3}{2}\pi$ in order to allow the EOS to cross $-1$ more than
one time within the redshift range from 0 to 2 where the SN data are
most robust, as implied by the recent studies \cite{Wu:2010av}.

Basing on the MCMC package CosmoMC\footnote{Available at:
http://cosmologist.info/cosmomc/.} \cite{CosmoMC} we perform a
global fitting analysis for the dynamical dark energy models
parameterized above. For dynamical dark energy models, it is crucial
to include dark energy perturbations
\cite{WMAP3GF,LewisPert,XiaPert}. As we know that for
quintessencelike or phantomlike models, in which $w$ does not cross
the cosmological constant boundary, the perturbation of dark energy
is well defined. However, when $w$ crosses $-1$, one is encountered
with the divergence problem for perturbations of dark energy at
$w=-1$. In order to solve this problem in the global fit analysis,
we introduce a small positive constant $\epsilon$ to divide the full
range of the allowed values of the EOS $w$ into three parts: (i) $ w
> -1 + \epsilon$, (ii) $-1 - \epsilon \leq w  \leq-1 + \epsilon$, and
(iii) $w < -1 -\epsilon $.

Working in the conformal Newtonian gauge, the perturbations of dark
energy can be described by \bea
    \dot\delta&=&-(1+w)(\theta-3\dot{\Phi})
    -3\mathcal{H}(c_{s}^2-w)\delta, \label{dotdelta}\\
\dot\theta&=&-\mathcal{H}(1-3w)\theta-\frac{\dot{w}}{1+w}\theta
    +k^{2}(\frac{c_{s}^2\delta}{{1+w}}+ \Psi). \label{dottheta}
\eea Neglecting the entropy perturbation, for the regions (i) and
(iii), the EOS does not cross $-1$ and the perturbation is well
defined by solving Eqs.~(\ref{dotdelta}) and (\ref{dottheta}). For
the case (ii), the perturbation of energy density $\delta$ and
divergence of velocity $\theta$, and the derivatives of $\delta$ and
$\theta$ are finite and continuous for the realistic dark energy
models. However for the perturbations of the parameterizations,
there is clearly a divergence. In our analysis for such a regime, we
match the perturbations in region (ii) to the regions (i) and (iii)
at the boundary and set $\dot{\delta}=0$ and $\dot{\theta}=0$. In
our numerical calculations we limit the range to be $|\Delta w =
\epsilon |<10^{-4}$ and find our method to be a very good
approximation to the multi-field dark energy model. More detailed
treatments can be found in Ref.~\cite{Zhao:2005vj}.

Our most general parameter space vector is:
\begin{equation}
\label{parameter} {\bf P} \equiv (\omega_{b}, \omega_{c},
\Theta, \tau, w_{0}, w_{1}, \Omega_k,  n_{s}, A_{s}, c_s^2),
\end{equation}
where $\omega_{b}\equiv\Omega_{b}h^{2}$ and
$\omega_{c}\equiv\Omega_{c}h^{2}$, with $\Omega_{b}$ and
$\Omega_{c}$ the physical baryon and cold dark matter densities
relative to the critical density, $\Omega_k$ is the spatial
curvature satisfying $\Omega_k+\Omega_b+\Omega_c+\Omega_{de}=1$,
$\Theta$ is the ratio (multiplied by 100) of the sound horizon
to the angular diameter distance at decoupling, $\tau$ is the
optical depth to re-ionization, $w_0$ and $w_1$ are the parameters
of dark energy EOS given by Eqs.~(\ref{pmt1}) and (\ref{pmt2}),
$A_s$ and $n_s$ are the amplitude and the spectral index of the
primordial scalar perturbation power spectrum, and $c_s$ is the
sound speed of dark energy. For the pivot scale we set
$k_{s0}=0.05$Mpc$^{-1}$. Note that we have assumed purely adiabatic
initial conditions.

In the computation of the CMB, we include the 7-year WMAP
temperature and polarization power spectra \cite{Komatsu:2010fb}
with the routine for computing the likelihood supplied by the WMAP
team.\footnote{Available at the LAMBDA website:
http://lambda.gsfc.nasa.gov/.} For the large scale structure (LSS)
information, we use the matter power spectrum data from SDSS DR7
\cite{Abazajian:2008wr}. The supernova data we use are the recently
released ``Union2'' sample of 557 data \cite{Riess:2011yx}. In the
calculation of the likelihood from SN we marginalize over the
relevant nuisance parameter \cite{SNMethod}.

Furthermore, we make use of the Hubble Space Telescope (HST)
measurement of the Hubble constant $H_{0}\equiv
100h$~km~s$^{-1}$~Mpc$^{-1}$ by a Gaussian likelihood function
centered around $h=0.738$ and with a standard deviation
$\sigma=0.024$ \cite{Riess:2011yx}.


\section{Numerical Results}

In this section we present our global fitting results. In
Table~\ref{tab1} we list the $1\sigma$ constraint results on the
dark energy models. We have compared the results with and without
the inclusion of the systematic errors of SN Union2 sample. By
including the systematic errors, the constraints on cosmological
parameters become a little bit relaxed, which can be seen by
comparing the error bars listed in the table for the two cases. Note
that the fit results listed in Table~\ref{tab1} are the mean of the
likelihood.

\begin{table*}\caption{Constraints on the dark energy EOS and some background parameters
from the observations.} 
\begin{center}\label{tab1}
\begin{tabular}{c|c|c|c|c|c}
\hline\hline
model & data&$\Omega_{de}$&$w_0$&$w_1$&$H_0$~(km~s$^{-1}$~Mpc$^{-1}$) \\
\hline
Log & ~Union2 (w/ sys)$+$WMAP7$+$LSS&$0.726^{+0.0204}_{-0.0207}$&$-0.951^{+0.0989}_{-0.100}$&$0.975^{+1.800}_{-1.864}$&$70.461^{+2.414}_{-2.429}$\\
\hline
Log & ~Union2 (w/o sys)$+$WMAP7$+$LSS&$0.729^{+0.0195}_{-0.0194}$&$-0.952^{+0.0911}_{-0.0923}$&$1.106^{+1.686}_{-1.744}$&$70.879^{+2.054}_{-2.064}$\\
\hline
Osc & ~Union2 (w/ sys)$+$WMAP7$+$LSS&$0.720^{+0.0120}_{-0.0124}$&$-0.959^{+0.0823}_{-0.0871}$&$0.0935^{+0.144}_{-0.143}$&$69.490^{+1.351}_{-1.381}$\\
\hline
Osc & ~Union2 (w/o sys)$+$WMAP7$+$LSS&$0.721^{+0.0119}_{-0.0120}$&$-0.964^{+0.0802}_{-0.0870}$&$0.0673^{+0.103}_{-0.107}$&$69.542^{+1.351}_{-1.343}$\\
\hline\hline
\end{tabular}
\end{center}
\end{table*}

Since our aim is to probe the dynamics of dark energy, we should try
to avoid other indirect factors weakening the observational limits
on the EOS. Thus, in our analysis we have assumed a flat universe,
$\Omega_k=0$, consistent with the inflationary cosmology. Moreover,
the sound speed of dark energy is also fixed in our analysis. In the
framework of the linear perturbation theory, besides the EOS of dark
energy, the dark energy perturbations can also be characterized by
the sound speed, $c_s^2\equiv\delta p_{de}/\delta\rho_{de}$. The
sound speed of dark energy might affect the evolution of
perturbations, and might leave signatures on the CMB power spectrum
\cite{Xia:2007km}. However, it has been shown that the constraints
on the dark energy sound speed $c_s^2$ in dynamical dark energy
models are still very weak, since the current observational data are
still not accurate enough \cite{Li:2010ac}. Therefore, in our
analysis, we have treated the dark energy as a scalar-field model
(multi-fields or single field with high derivative) and set $c_s^2$
to be 1. Of course, one can also take $c_s^2$ as a parameter, but
the fit results would not be affected by this treatment
\cite{Li:2010ac}.

\begin{figure*}[htbp]
\centering
\begin{center}
$\begin{array}{c@{\hspace{0.2in}}c} \multicolumn{1}{l}{\mbox{}} &
\multicolumn{1}{l}{\mbox{}} \\
\includegraphics[scale=0.4]{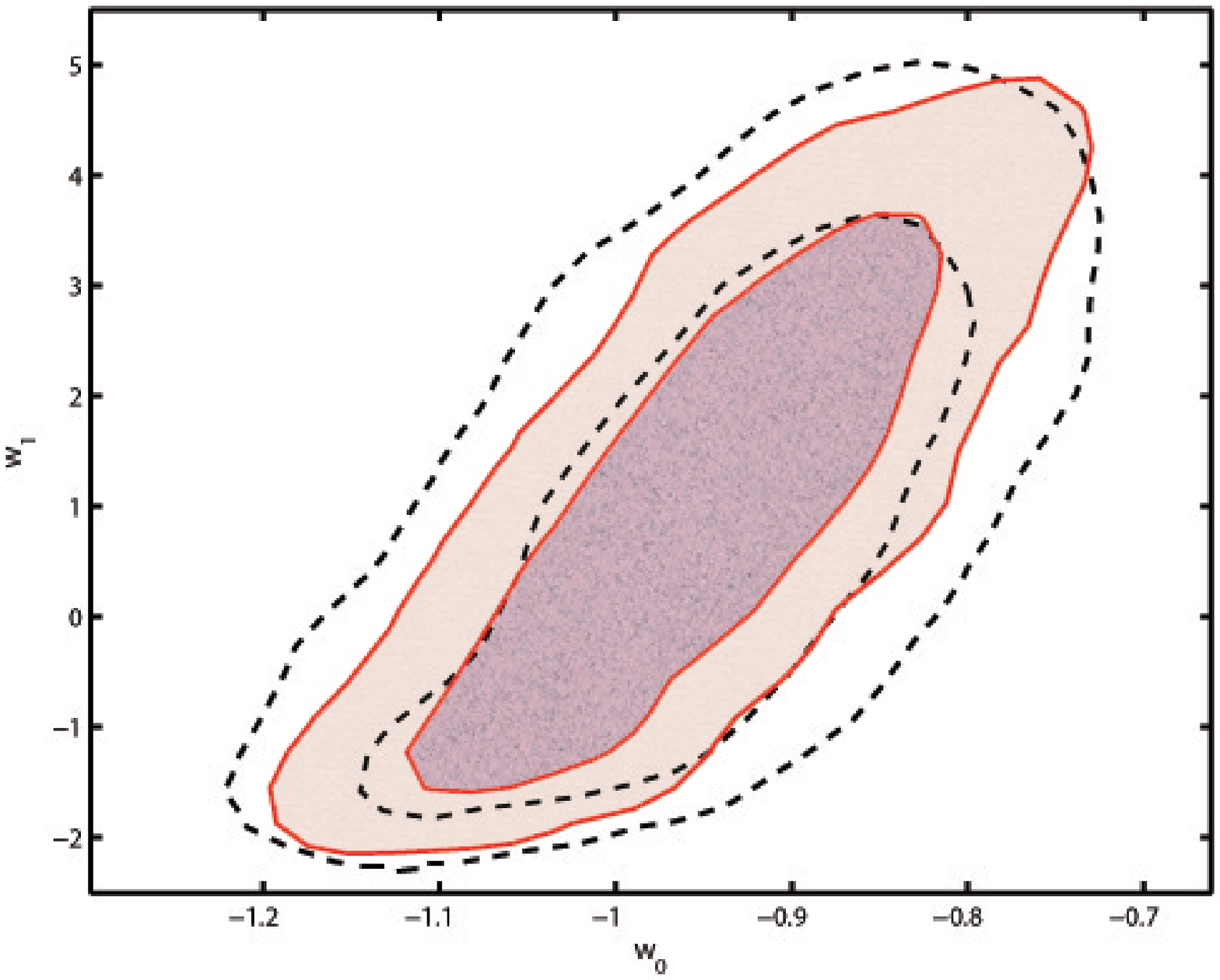}&\includegraphics[scale=0.4]{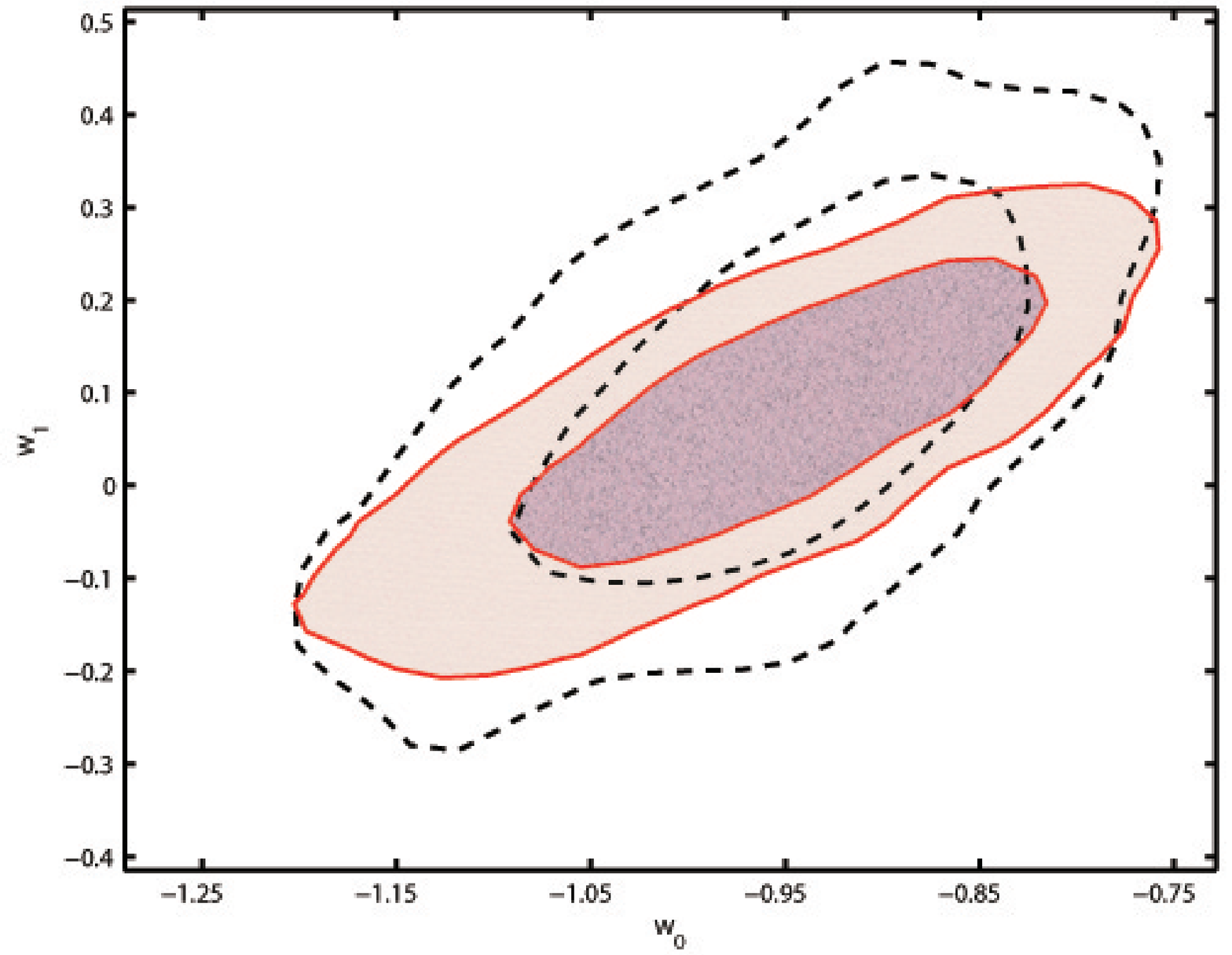} \\
\mbox{(a)} & \mbox{(b)}
\end{array}$
\end{center}
\caption[]{\small Joint two-dimensional marginalized constraint on
the parameters $w_0$ and $w_1$ for (a) the logarithm parametrization
(\ref{pmt1}) and (b) the oscillating parametrization (\ref{pmt2}).
The contours show the 68$\%$ and 95$\%$ CL from WMAP+SDSS+SN, for
the cases without the systematic errors of SN (color shaded regions
and red solid lines) and with the systematic errors of SN (unshaded
regions and black dashed lines). \label{fig1}}
\end{figure*}

\begin{figure*}[htbp]
\centering
\begin{center}
$\begin{array}{c@{\hspace{0.2in}}c} \multicolumn{1}{l}{\mbox{}} &
\multicolumn{1}{l}{\mbox{}} \\
\includegraphics[scale=0.4]{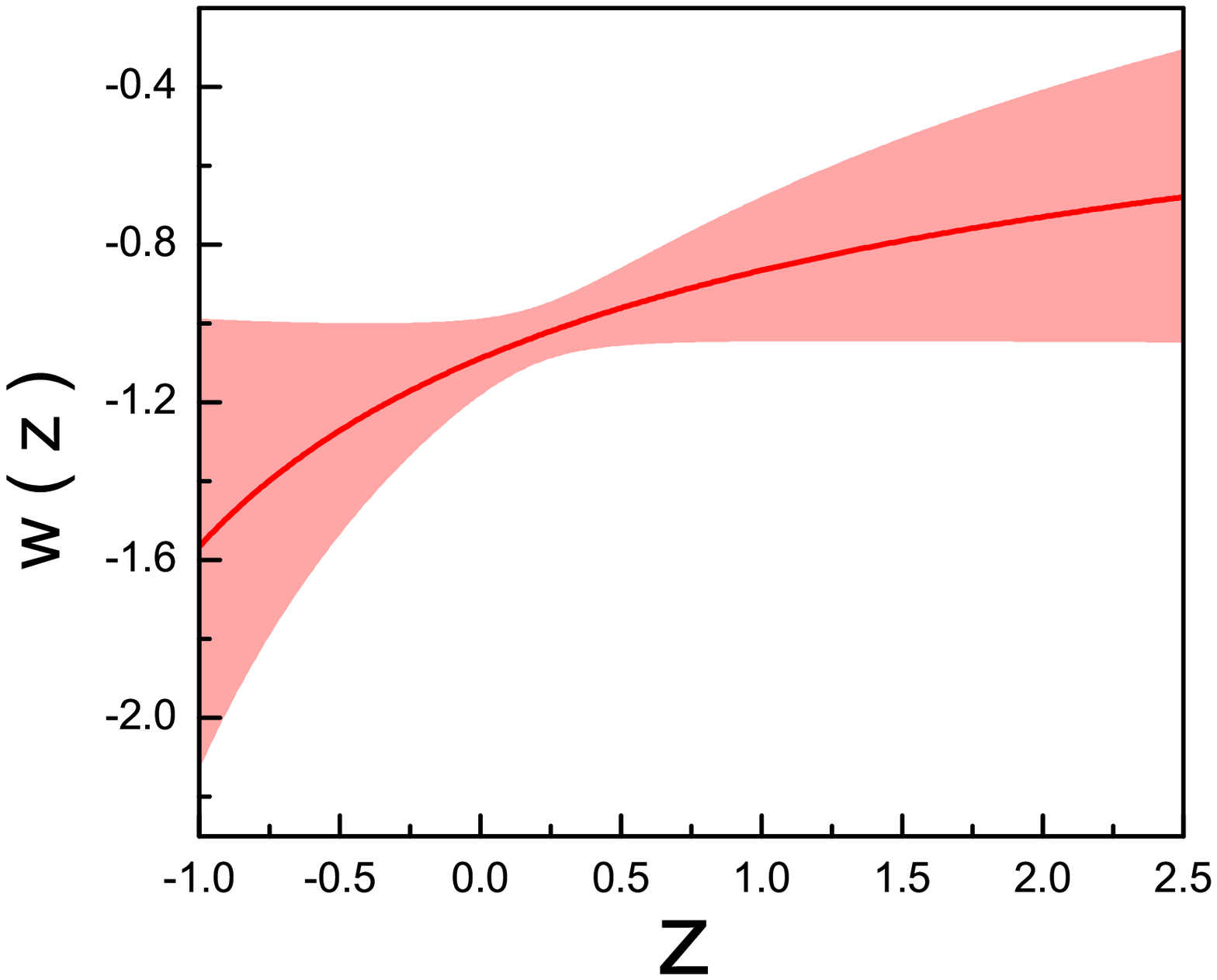} &\includegraphics[scale=0.4]{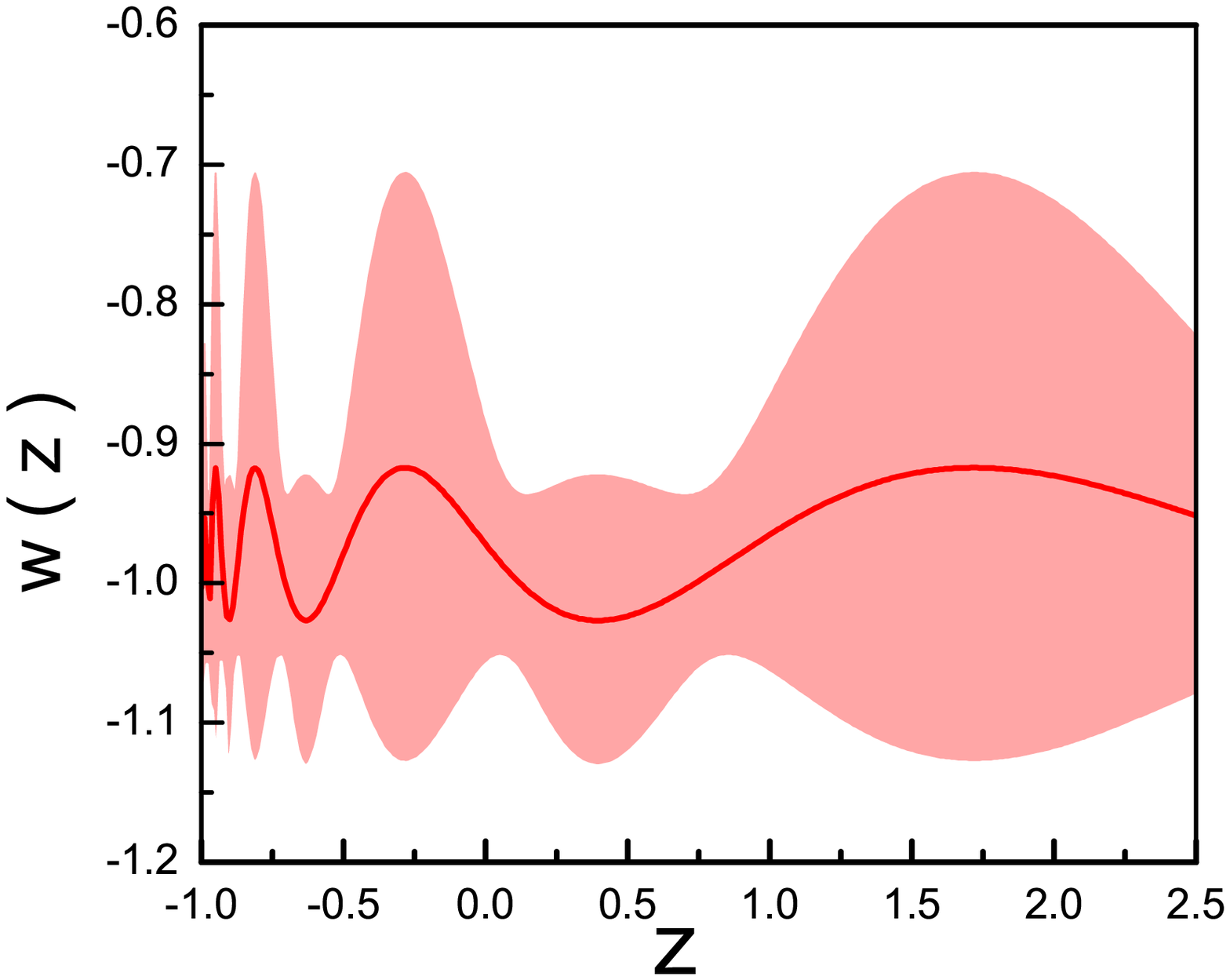} \\
\mbox{(a)} & \mbox{(b)}
\end{array}$
\end{center}
\caption[]{\small The reconstructed evolutionary histories (from
past to future) for $w(z)$ of the two dark energy models: (a) the
logarithm model and (b) the oscillating model. The red solid line is
plotted with the best fit values, while the color shaded region is
given by the $1\sigma$ limit. \label{fig2}}
\end{figure*}

In the dynamical dark energy model where the EOS of dark energy is
parameterized by the logarithm form (\ref{pmt1}), we get the
best-fit results (including the systematic errors of Union2 sample):
$\omega_{b}=0.0225$, $\omega_{c}=0.114$, $h=0.683$, $\tau=0.0877$,
$w_{0}=-1.089$, $w_{1}=-1.552$, $n_{s}=0.969$ and
$10^{9}A_{s}=2.207$, which are consistent with the results of 7-year
WMAP \cite{Komatsu:2010fb}. Note that here the results are maximum
likelihood values. The panel (a) of Fig.~\ref{fig1} shows the joint
two-dimensional marginalized constraint on the parameters $w_0$ and
$w_1$ for the logarithm parametrization (\ref{pmt1}). The contours
show the 68$\%$ and 95$\%$ confidence levels (CL) for the cases
without the systematic errors of SN (color shaded regions and red
solid lines) and with the systematic errors of SN (unshaded regions
and black dashed lines). We find that the best-fit dark energy model
is a quintom model \cite{quintom}, whose $w(z)$ crosses the
cosmological constant boundary $w=-1$ during the evolution. With the
current observational data, the variance of $w_0$ and $w_1$ we get
are still large; the $95\%$ constraints on $w_0$ and $w_1$ are
$-1.154<w_0<0.771$ and $-1.682<w_1<4.251$, which can also be seen in
the panel (a) of Fig.\ref{fig1}. This result implies that though the
dynamical dark energy models are mildly favored, the current data
cannot distinguish different dark energy models decisively. With the
fitting results in hand, we can reconstruct the evolution of the EOS
of dark energy, $w(z)$. The reconstructed result for the logarithm
form (\ref{pmt1}) is shown in the panel (a) of Fig.~\ref{fig2}. The
red solid line is plotted with the best fit values, while the color
shaded region represents the $1\sigma$ limit. From this figure, we
can directly see that although the quintom model is more favored,
the cosmological constant ($\Lambda$CDM model), however, still
cannot be excluded.

Our results are consistent with Ref.~\cite{Ma:2011nc}, in which the
constraints are given by using the data combination of the WMAP
distance prior and BAO information instead of the full CMB
temperature and polarization power spectra and LSS matter power
spectrum. Though the WMAP distance prior, including $R$, $l_A$ and
$z_{*}$, encoding the information of background cosmic distances,
can be applied to investigate dark energy models and can greatly
simplify the numerical calculations in determining cosmological
parameters, it was found that the prior is somewhat cosmological
model dependent and the utilization of this prior may lose some of
the CMB information \cite{Komatsu:2010fb}. For example, the distance
prior does not capture the information on the growth of structure
probed by the late-time ISW effect. As a result, the dark energy
constraints derived from the distance prior are similar to, but
weaken than, those derived from the full analysis. Therefore, in
this Letter, in order to improve the analysis in
Ref.~\cite{Ma:2011nc}, we present the full Markov Chain Monte Carlo
exploration of this model.

Next, we discuss the dynamical dark energy model with the EOS of
dark energy parameterized by the oscillatory form (\ref{pmt2}). For
this model, we get the best-fit results: $\omega_{b}=0.0225$,
$\omega_{c}=0.112$, $h=0.697$, $\tau=0.0878$, $w_{0}=-1.089$,
$w_{1}=-1.553$,  $n_{s}=0.969$ and $10^{9}A_{s}=2.207$. We find that
the EOS of dark energy that has an oscillating behavior can also fit
the data well. The $2\sigma$ CL constraints on $w_0$ and $w_1$ in
this dark energy model are $-1.149<w_0<-0.810$ and
$-0.192<w_1<0.357$. We show the two-dimensional marginalized
constraint on $w_0$ and $w_1$ for this model in the panel (b) of
Fig.~\ref{fig1}. The reconstructed evolution behavior of the EOS of
dark energy, $w(z)$, is shown in the panel (b) of Fig.~\ref{fig2}.
It is indicated that an oscillating quintom model is more favored,
whose EOS crosses $-1$ more than one time within the redshift range
from 0 to 2. According to the best-fit result, for the future
evolution, the EOS of this model will experience the $-1$ crossing
for many times. However, one can also see that at the $1\sigma$
level, the cosmological constant is still a good fit.


\section{Summary and discussion}
\label{Sum}

In this Letter, we have performed a global fit study on two
divergence-free parametrizations for dark energy. It is known that
the frequently used CPL parametrization actually can only describe
the past evolution history of dark energy but cannot genuinely
depict the future evolution of dark energy owing to the divergence
of $w(z)$ as $z$ approaches $-1$. Such a divergence feature forces
the CPL parametrization to lose its prediction capability for the
fate of the universe and to fail in providing a complete evolution
history for the dark energy. Consequently, the CPL model cannot
genuinely cover scalar-field models as well as other dark energy
theoretical models. In Ref.~\cite{Ma:2011nc}, a robust
parametrization form, $w(z)=w_0+w_1({\ln (2+z)\over 1+z}-\ln2)$, was
proposed, which is divergence-free and has well-behaved feature for
the EOS of dark energy in all the evolution stages of the universe.

This parametrization, without doubt, could cover many dark-energy
theoretical models. Obviously, according to this parametrization,
quintom models with two scalar fields and/or with one field with
high derivatives can be successfully reconstructed. Another example
can be provided by the holographic dark energy model
\cite{Limiao:2004rb}. The holographic dark energy model arises from
the holographic principle of quantum gravity. Its EOS satisfies
$w(z)=-1/3-2/(3c)\sqrt{\Omega_{de}(z)}$, where $c$ is a
phenomenological parameter determining the dynamical evolution of
the dark energy, and $\Omega_{de}(z)$ satisfies a differential
equation \cite{holofitzx}. In this model, if $c<1$, the dark energy
will behave like a quintom, i.e., the EOS crosses $-1$ during the
evolution \cite{Zhang:2005yz}. We find that the holographic
evolution can be roughly mimicked by the logarithm form
parametrization, provided that $w_0$ and $w_1$ are around $-1$. All
in all, this parametrization can be used to reconstruct many
dark-energy theoretical models, and can be used to probe the
dynamics of dark energy in light of the observational data.

In Ref.~\cite{Ma:2011nc}, the logarithm parametrization (\ref{pmt1})
has been used to probe the dynamics of dark energy in the whole
evolutionary history. However, it should be pointed out that only a
preliminary analysis was performed in Ref.~\cite{Ma:2011nc} because
the data sets in the analysis are the WMAP distance prior and BAO
information instead of the full CMB temperature and polarization
power spectra and LSS matter power spectrum. Such an analysis might
lose some of the CMB and LSS information. In this Letter, we have
improved the analysis by implementing a full Markov Chain Monte
Carlo exploration of this model. The result is consistent with that
of Ref.~\cite{Ma:2011nc}. We found that the best-fit dark energy
model is a quintom model with the EOS $w(z)$ across $-1$ during the
evolution. However, while the quintom model is more favored, the
cosmological constant still cannot be excluded.

We also explored the possibility that the EOS may oscillate and
cross $-1$ many times during the evolution. We used the oscillating
parametrization (\ref{pmt2}) that is also divergence-free to probe
the evolution of dark energy. Though the motivation of this
parametrization is not as robust as the form (\ref{pmt1}), it can
fit the data well. The result shows that it is indeed possible that
$w(z)$ crosses $-1$ for many times during the whole evolution
history.

We believe that it is fairly important to use some divergence-free
parametrization forms to probe the dynamical evolution of dark
energy. We have shown that the logarithm form (\ref{pmt1}) is a good
proposal and it has been proven to be very successful in exploring
the property of dark energy. We suggest that this parametrization
should be further investigated.


\section*{Acknowledgements}

We acknowledge the use of the Legacy Archive for Microwave
Background Data Analysis (LAMBDA). Support for LAMBDA is provided by
the NASA Office of Space Science. The calculation is taken  on
Deepcomp7000 of Supercomputing Center, Computer Network Information
Center of Chinese Academy of Sciences.  This work is supported in
part by the National Science Foundation of China under Grant
Nos.~11033005, 10975032, 10803001, and 10705041, by the 973 program
under Grant No.~2010CB83300, by the National Ministry of Education
of China under Grant No. NCET-09-0276, and by the Youth Foundation
of the Institute of High Energy Physics under Grant No. H95461N.



\begin{thebibliography}{nn}

\bibitem{Riess98}
  A.~G.~Riess {\it et al.}  [Supernova Search Team Collaboration],
  Astron.\ J.\  {\bf 116}, 1009 (1998)
  [astro-ph/9805201];
  S.~Perlmutter {\it et al.}  [Supernova Cosmology Project Collaboration],
  Astrophys.\ J.\  {\bf 517}, 565 (1999)
  [astro-ph/9812133].

\bibitem{dereview}
  V.~Sahni and A.~A.~Starobinsky,
  Int.\ J.\ Mod.\ Phys.\  D {\bf 9}, 373 (2000)
  [arXiv:astro-ph/9904398];
  P.~J.~E.~Peebles and B.~Ratra,
  Rev.\ Mod.\ Phys.\  {\bf 75}, 559 (2003)
  [arXiv:astro-ph/0207347];
  V.~Sahni,
  Lect.\ Notes Phys.\  {\bf 653}, 141 (2004)
  [arXiv:astro-ph/0403324];
  E.~J.~Copeland, M.~Sami and S.~Tsujikawa,
  Int.\ J.\ Mod.\ Phys.\  D {\bf 15}, 1753 (2006)
  [arXiv:hep-th/0603057];
  J.~Frieman, M.~Turner and D.~Huterer,
  Ann.\ Rev.\ Astron.\ Astrophys.\  {\bf 46}, 385 (2008)
  [arXiv:0803.0982 [astro-ph]];
  M.~Li, X.~D.~Li, S.~Wang and Y.~Wang,
  arXiv:1103.5870 [astro-ph.CO].


\bibitem{de2}
  H.~Li, M.~Su, Z.~Fan, Z.~Dai and X.~M.~Zhang,
  Phys.\ Lett.\  B {\bf 658}, 95 (2008)
  [arXiv:astro-ph/0612060];
  J.~Q.~Xia, H.~Li, G.~B.~Zhao and X.~M.~Zhang,
  Phys.\ Rev.\  D {\bf 78}, 083524 (2008)
  [arXiv:0807.3878 [astro-ph]];
  H.~Li {\it et al.},
  Phys.\ Lett.\  B {\bf 675}, 164 (2009)
  [arXiv:0812.1672 [astro-ph]];
  M.~Li, X.~D.~Li and X.~Zhang,
  Sci.\ China Phys.\ Mech.\ Astron.\  {\bf 53}, 1631 (2010)
  [arXiv:0912.3988 [astro-ph.CO]];
  Q.~G.~Huang, M.~Li, X.~D.~Li and S.~Wang,
  Phys.\ Rev.\  D {\bf 80}, 083515 (2009)
  [arXiv:0905.0797 [astro-ph.CO]];
  M.~Li, X.~D.~Li, S.~Wang, Y.~Wang and X.~Zhang,
  JCAP {\bf 0912}, 014 (2009)
  [arXiv:0910.3855 [astro-ph.CO]];
  S.~Wang, X.~D.~Li and M.~Li,
  Phys.\ Rev.\  D {\bf 83}, 023010 (2011)
  [arXiv:1009.5837 [astro-ph.CO]];
  X.~D.~Li, S.~Li, S.~Wang, W.~S.~Zhang, Q.~G.~Huang and M.~Li,
  JCAP {\bf 1107}, 011 (2011)
  [arXiv:1106.4116 [astro-ph.CO]].







\bibitem{CPL}
 M.~Chevallier and D.~Polarski,
 Int.\ J.\ Mod.\ Phys.\ D {\bf 10}, 213 (2001)
  [arXiv:gr-qc/0009008];
 E.~V.~Linder,
 Phys.\ Rev.\ Lett.\  {\bf 90}, 091301 (2003)
 [arXiv:astro-ph/0208512].

\bibitem{Ma:2011nc}
  J.~Z.~Ma and X.~Zhang,
  Phys.\ Lett.\  B {\bf 699}, 233 (2011)
  [arXiv:1102.2671 [astro-ph.CO]].


\bibitem{Dodelson:2001fq}
  S.~Dodelson, M.~Kaplinghat and E.~Stewart,
  Phys.\ Rev.\ Lett.\  {\bf 85}, 5276 (2000)
  [arXiv:astro-ph/0002360].

\bibitem{Feng:2004ff}
  B.~Feng, M.~Li, Y.~S.~Piao and X.~M.~Zhang,
  Phys.\ Lett.\  B {\bf 634}, 101 (2006)
  [arXiv:astro-ph/0407432].

\bibitem{Xia:2004rw}
  J.~Q.~Xia, B.~Feng and X.~M.~Zhang,
  Mod.\ Phys.\ Lett.\  A {\bf 20}, 2409 (2005)
  [arXiv:astro-ph/0411501].

\bibitem{Linder:2005dw}
  E.~V.~Linder,
  Astropart.\ Phys.\  {\bf 25}, 167 (2006)
  [arXiv:astro-ph/0511415].

\bibitem{Zhao:2006qg}
  G.~B.~Zhao, J.~Q.~Xia, H.~Li, C.~Tao, J.~M.~Virey, Z.~H.~Zhu and X.~M.~Zhang,
  Phys.\ Lett.\  B {\bf 648}, 8 (2007)
  [arXiv:astro-ph/0612728].

\bibitem{Wu:2010av}
  P.~Wu and H.~W.~Yu,
  Eur.\ Phys.\ J.\  C {\bf 71}, 1552 (2011)
  [arXiv:1008.3669 [gr-qc]];
  K.~Bamba, C.~Q.~Geng, C.~C.~Lee and L.~W.~Luo,
  JCAP {\bf 1101}, 021 (2011)
  [arXiv:1011.0508 [astro-ph.CO]].







   \bibitem{CosmoMC}
  A.~Lewis and S.~Bridle,
  Phys.\ Rev.\  D {\bf 66}, 103511 (2002).

  \bibitem{WMAP3GF}
D.~Spergel, \emph{et al.}, Astrophys.\ J.\ Suppl.\ 170 (2007) 377.

 \bibitem{LewisPert}
J.~Weller and A.~Lewis, Mon.\ Not.\ R.\ Astron.\ Soc.\ 346 (2003)
987.

\bibitem{XiaPert}
J.~Q.~Xia, G.~B.~Zhao, B.~Feng, H.~Li, X.~M.~Zhang, Phys.\ Rev.\ D
73 (2006) 063521.

\bibitem{Zhao:2005vj}
  G.~B.~Zhao, J.~Q.~Xia, M.~Li, B.~Feng and X.~M.~Zhang,
  Phys.\ Rev.\  D {\bf 72}, 123515 (2005)
  [arXiv:astro-ph/0507482].

\bibitem{Komatsu:2010fb}
  E.~Komatsu {\it et al.}  [WMAP Collaboration],
  Astrophys.\ J.\ Suppl.\  {\bf 192}, 18 (2011)
  [arXiv:1001.4538 [astro-ph.CO]].

\bibitem{Abazajian:2008wr}
  K.~N.~Abazajian {\it et al.}  [SDSS Collaboration],
  Astrophys.\ J.\ Suppl.\  {\bf 182}, 543 (2009)
  [arXiv:0812.0649 [astro-ph]].


\bibitem{Riess:2011yx}
  A.~G.~Riess {\it et al.},
  Astrophys.\ J.\  {\bf 730}, 119 (2011)
  [Erratum-ibid.\  {\bf 732}, 129 (2011)]
  [arXiv:1103.2976 [astro-ph.CO]].

   \bibitem{SNMethod}
E.~Di~Pietro and J.~F.~Claeskens, Mon.\ Not.\ Roy.\ Astron.\ Soc.\
341, 1299 (2003).

\bibitem{Xia:2007km}
  J.~Q.~Xia, Y.~F.~Cai, T.~T.~Qiu, G.~B.~Zhao and X.~Zhang,
  Int.\ J.\ Mod.\ Phys.\  D {\bf 17}, 1229 (2008)
  [arXiv:astro-ph/0703202].

\bibitem{Li:2010ac}
  H.~Li and J.~Q.~Xia,
  JCAP {\bf 1004}, 026 (2010)
  [arXiv:1004.2774 [astro-ph.CO]].

\bibitem{quintom}
  B.~Feng, X.~-L.~Wang, X.~-M.~Zhang,
  Phys.\ Lett.\  B {\bf 607}, 35 (2005)
  [astro-ph/0404224];
  Z.~K.~Guo, Y.~S.~Piao, X.~M.~Zhang and Y.~Z.~Zhang,
  Phys.\ Lett.\  B {\bf 608}, 177 (2005)
  [astro-ph/0410654];
  X.~Zhang,
  Commun.\ Theor.\ Phys.\  {\bf 44}, 762 (2005);
  X.~F.~Zhang, H.~Li, Y.~S.~Piao and X.~M.~Zhang,
  Mod.\ Phys.\ Lett.\  A {\bf 21}, 231 (2006)
  [arXiv:astro-ph/0501652];
  Y.~-F.~Cai, E.~N.~Saridakis, M.~R.~Setare, J.~-Q.~Xia,
  Phys.\ Rept.\  {\bf 493}, 1 (2010)
  [0909.2776 [hep-th]].

\bibitem{Limiao:2004rb}
  M.~Li,
  Phys.\ Lett.\  B {\bf 603}, 1 (2004)
  [arXiv:hep-th/0403127].

\bibitem{holofitzx}
  Q.~G.~Huang and Y.~G.~Gong,
  JCAP {\bf 0408}, 006 (2004)
  [arXiv:astro-ph/0403590];
  X.~Zhang and F.~Q.~Wu,
  Phys.\ Rev.\  D {\bf 72}, 043524 (2005)
  [arXiv:astro-ph/0506310];
  X.~Zhang and F.~Q.~Wu,
  Phys.\ Rev.\  D {\bf 76}, 023502 (2007)
  [arXiv:astro-ph/0701405];
  Z.~Chang, F.~Q.~Wu and X.~Zhang,
  Phys.\ Lett.\  B {\bf 633}, 14 (2006)
  [arXiv:astro-ph/0509531];
  M.~Li, X.~D.~Li, S.~Wang and X.~Zhang,
  JCAP {\bf 0906}, 036 (2009)
  [arXiv:0904.0928 [astro-ph.CO]].



\bibitem{Zhang:2005yz}
  X.~Zhang,
  Int.\ J.\ Mod.\ Phys.\  D {\bf 14}, 1597 (2005)
  [arXiv:astro-ph/0504586];
  X.~Zhang,
  Phys.\ Rev.\  D {\bf 74}, 103505 (2006)
  [arXiv:astro-ph/0609699];
  X.~Zhang,
  Phys.\ Lett.\  B {\bf 648}, 1 (2007)
  [arXiv:astro-ph/0604484];
  J.~Zhang, X.~Zhang and H.~Liu,
  Phys.\ Lett.\  B {\bf 651}, 84 (2007)
  [arXiv:0706.1185 [astro-ph]];
  J.~Zhang, X.~Zhang and H.~Liu,
  Eur.\ Phys.\ J.\  C {\bf 52}, 693 (2007)
  [arXiv:0708.3121 [hep-th]];
  X.~Zhang,
  Phys.\ Lett.\  B {\bf 683}, 81 (2010)
  [arXiv:0909.4940 [gr-qc]].








\end{thebibliography}
\end{document}